# 3D Polarization-Independent Visible Invisibility

Joachim Fischer, Tolga Ergin, and Martin Wegener

*Institut für Angewandte Physik, Institut für Nanotechnologie, and DFG-Center for Functional Nanostructures (CFN), Karlsruhe Institute of Technology (KIT), D-76128 Karlsruhe, Germany*

[joachim.fischer@kit.edu](joachim.fischer@kit.edu)

"Seeing is believing". Invisibility cloaking appeals to laymen and scientists alike because a phenomenon long believed to be impossible has suddenly come into reach. However, the holy grail of a three-dimensional (3D) cloak for visible unpolarized light has not been accomplished so far. Direct laser writing optical lithography (DLW), that has been successfully used to fabricate 3D structures (*1*), "just" has to be improved by a bit more than a factor of two in *all* spatial directions to achieve that goal. Yet, improving optical lithography already operated at the diffraction limit by that factor is not a minor task at all.

Here, we accomplish this step by employing stimulated-emission-depletion inspired DLW (STED-DLW), with conceptually diffraction-unlimited axial and lateral resolution.

By transferring concepts from stimulated-emission-depletion (STED) fluorescence microscopy (*2*) to DLW, the diffraction barrier has recently started to erode (*3-5*). In essence, one laser serves as a pencil of light in three dimensions and a second laser serves as an eraser, allowing for "erasing" the edges of written lines, hence reducing the feature sizes. However, transferring this advance to complex free-standing 3D nanostructures has not been demonstrated so far. Furthermore, it is most crucial to improve the axial resolution in DLW, since it has been nearly three times worse than the lateral resolution for relevant lens numerical apertures of *NA*=1.4, hence limiting the *overall* resolution of 3D DLW.

Our approach is illustrated in Fig. 1A. As usual, optical femtosecond pulses centered at around 810-nm wavelength are tightly focused by a microscope objective with *NA*=1.4 to polymerize a photoresist within the focal volume *via* two-photon absorption. By using a home-made optical phase-mask (see SOM), a continuous-wave 532-nm wavelength "eraser" beam is shaped such that the elongated red focal volume is effectively reduced to the desired more spherical exposure volume. We have designed a photoresist (see SOM) following our spirit (*5*) that the anticipated stimulated emission depletion is only possible for reasonably efficient spontaneous emission. In contrast to our previous choice (*5*), the photoinitiator used here offers a better ratio between depletion efficiency and undesired absorption of the continuous-wave depletion laser (which limits resolution). Importantly, we presently see no fundamental optical limitation of our approach in terms of obtainable minimum feature sizes in three dimensions – just like in STED microscopy.

We have fabricated a number of carpet cloak structures with this type of optical lithography. In a nutshell, the carpet cloak (*1,6*) bends light in such a way that a bump in a metallic carpet appears to be flat. Hence, objects hidden underneath the bump become invisible. To do so, the local phase velocity of light (thus the refractive index) has to be tailored following the mathematics of transformation optics. The required refractive index distribution can be realized by adjusting the local volume filling fraction of a woodpile photonic crystal. In our 2010 experimental work (*1*), we have succeeded in fabricating such a device with a rod spacing of *a*=800 nm. Here, to allow for direct comparison, we fabricate a miniaturized version of the 3D carpet cloak with *a*=350 nm. Due to the scalability of the Maxwell equations, the operation wavelength is expected to scale from the lower limit of about 1.5 µm in Ref.(*1*) down to
1.5 µm × 350/800 = 0.7 µm, *i.e.*, to visible red light. Corresponding structures, control experiments, and successful cloaking results are shown in Fig. 1B, C, and D, respectively.

In a broader context, the special example of a polarization-independent invisibility cloak at visible wavelengths demonstrated here raises hopes that diffraction-unlimited optical lithography may truly become the 3D analogue of 2D electron-beam lithography, which has served as a workhorse for the entire field of (planar) nanotechnology for many years.

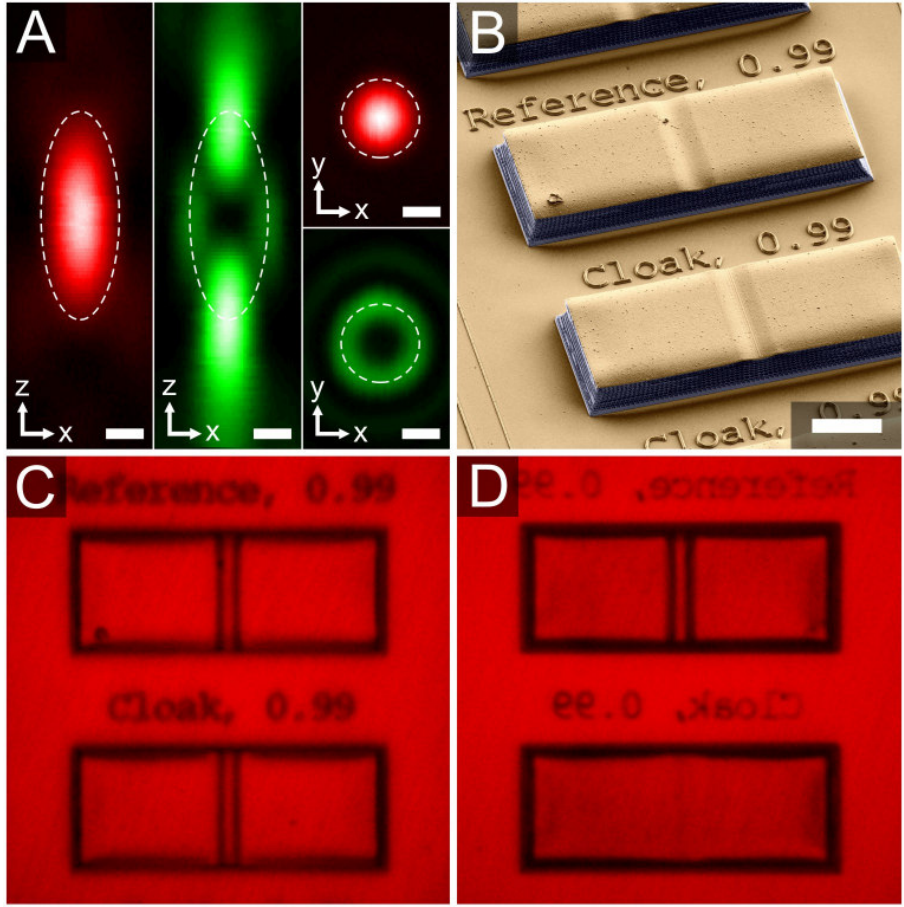

**Fig. 1. (A)** Measured focal intensity distributions at the 810-nm excitation (red) and 532-nm depletion (green) wavelengths in the axial *xz*-plane and the lateral *xy*-plane. The scale bars correspond to 250 nm. **(B)** Colored oblique-view electron micrograph of the polymer reference (top) and cloaking (bottom) structures (fabricated on glass substrate and coated with 100 nm gold). The scale bar corresponds to 10 µm. **(C)** and **(D)** are true-color optical micrographs of the structures in **(B)** taken with a microscope (ZEISS objective LD Achroplan, 20×, *NA*=0.4) under circularly polarized monochromatic illumination at 700-nm wavelength. **(C)** is a control experiment taken from the air side. Note the identical distortions due to the bump. **(D)** When inspected through the glass substrate, the reference structure (top) still shows pronounced dark stripes. In sharp contrast, the stripes disappear for the cloaking structure (bottom).

## Supporting Online Material

Our home-made photoresist consists of pentaerythritol tetraacrylate (Sigma Aldrich) and 0.25% wt 7-diethylamino-3-thenoylcoumarin (Exciton). The former contains 350 ppm monomethyl ether hydroquinone as inhibitor. The molar extinction coefficient of the photoinitiator is 35,000 l mol$^{-1}$ cm$^{-1}$ (Exciton data sheet), which is comparable to state-of-the-art dyes, *e.g.*, Atto425 with 45,000 l mol$^{-1}$ cm$^{-1}$ (ATTO-TEC GmbH). The cross-section for stimulated emission is proportional to these values. Measured absorbance and photoluminescence spectra of our photoinitiator in ethanol solution are shown in Fig. S1. This figure also shows the measured depletion of photoluminescence with increasing power of the depletion laser (without phase mask, *i.e.*, for a Gaussian depletion focus). This observation is consistent with stimulated emission being the underlying depletion mechanism. However, other mechanisms leading to photoinduced reversible polymerization suppression could contribute as well and are also expected to lead to axial and lateral resolution improvement. Hence, they serve the spirit of STED-DLW. After exposure, the photoresist structures are developed in 2-propanol and dried super-critically with $CO_2$ (Leica EM CPD030).

The special depletion focus in Fig. 1A is accomplished by a home-made phase mask. It consists of a 430-nm high cylinder (photoresist SU-8, MicroChem) with refractive index $n$=1.62 at 532-nm wavelength on a glass substrate. This arrangement introduces a 180-degrees phase shift in the center of the continuous-wave circularly polarized collimated beam. The phase-mask plane is imaged onto the entrance pupil of the microscope objective (Leica HCX PL APO, *NA*=1.4) of the DLW system such that the central area occupies about 50 % of the entrance pupil area. The focal intensity distributions of the red (Spectra Physics, Mai Tai HP) and green (Spectra Physics, Millennia Xs) beams in Fig. 1A are measured by scanning a single 100-nm diameter gold bead through the focus and recording the back-scattered light intensity. These intensity distributions show that the green depletion profile not only reduces the red profile along the axial direction, but also comprises a ring in the focal plane, improving the lateral resolution.

In the fabrication process, the femtosecond excitation (10-mW average power) and continuous-wave depletion (50 mW) lasers are chopped at 3% duty cycle (4-kHz frequency). The local filling fraction of the woodpile photonic crystal is controlled by varying the excitation power.

To enhance the visibility of details in Figure 1B, we have overlaid two images with different brightness (high dynamic range (HDR) image). The gold parts were colored in yellow, the polymer parts in blue.

The oblique-view (54-degrees tilt with respect to surface normal) electron micrograph of a focused-ion-beam cut of a cloaking structure shown in Fig. S1 reveals the cloak's interior. The woodpile rod spacing is $a$=350 nm. This value has never been achieved before by DLW *without* depletion beam. As a control experiment, we have switched off the green depletion laser, while keeping all other parameters fixed. Under these conditions, we have not even found any open polymer nanostructure for any power of the red excitation beam. The overall design dimensions of the cloaking structure in Fig. 1B are 50 µm × 20 µm × 5 µm. The full width of the cos²-shaped bump is 6 µm, its height is 0.5 µm. After completing optical lithography, the samples are coated with a 100-nm thick gold film in a sputter chamber (Cressington 108 auto).

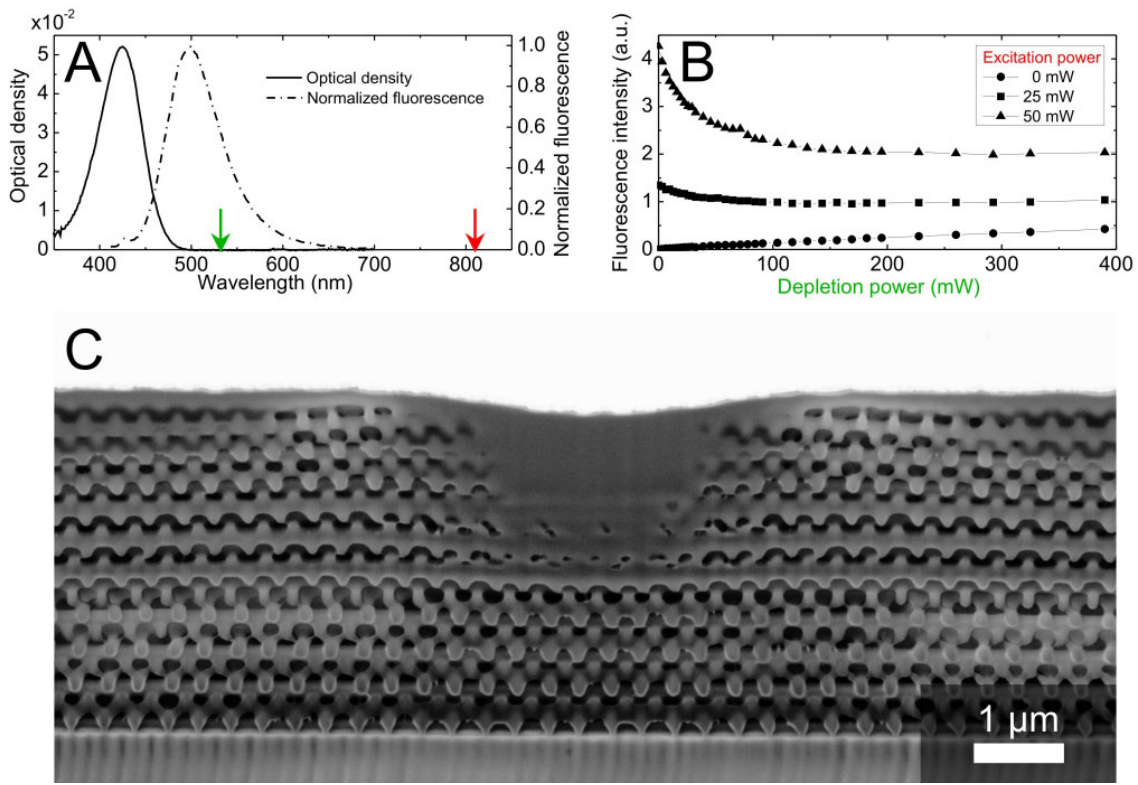

**Fig. S1**. (**A**) Optical density (absorbance) and fluorescence spectra of the photoinitiator in ethanol solution. The arrows mark the operation wavelengths of the lasers. (**B**) Depletion of the photoinitiator fluorescence with increasing depletion power. (**C**) Oblique-view electron micrograph of an *a*=350 nm cloaking structure made by STED-DLW after focused-ion-beam milling.